\documentclass[12pt]{article}
\usepackage{graphicx}
\usepackage{epsfig}
\usepackage{epsf}
\usepackage{amssymb}
\usepackage{rotate}
\newcommand{\ba}{\begin{array}}
\newcommand{\ea}{\end{array}}
\newcommand{\bd}{\begin{displaymath}}
\newcommand{\ed}{\end{displaymath}}
\newcommand{\be}{\begin{equation}}
\newcommand{\ee}{\end{equation}}
\newcommand{\bea}{\begin{eqnarray}}
\newcommand{\eea}{\end{eqnarray}}

\def\bra{\langle}
\def\ket{\rangle}

\def\a{\alpha}
\def\b{\beta}

\def\e{\epsilon}

\def\m{\mu}
\def\n{\nu}

\def\n{\nu}

\def\th13 {\theta_{13}}

\begin{document}
\thispagestyle{empty}
\begin{flushright}
\texttt{}\\

\end{flushright}
\vskip 15pt
\begin{center}
{\Large {\bf Light neutrinos from massless texture and below TeV
seesaw scale }}       \\
\vskip 30pt
\renewcommand{\thefootnote}{\alph{footnote}}
\large{{
Rathin Adhikari $\footnote{E-mail address:
rathin@ctp-jamia.res.in}^{\dagger}$,
}}\\
\large{{ and }}
\\
\large{{ Amitava Raychaudhuri
$\footnote{E-mail address: raychaud@hri.res.in}^{\ddagger \;\star}$}}\\
\vskip 20pt
\small$\phantom{i}^{\dagger}${\em Centre for Theoretical Physics,\\
 Jamia Millia
Islamia (Central University), New Delhi - 110025, India}\\

\vskip 10pt
\small$\phantom{i}^{\ddagger}${\em Harish-Chandra Research Institute,\\
Chhatnag Road, Jhunsi, Allahabad - 211019, India }   \\

\vskip 10 pt

\small$\phantom{i}^{\star}${\em Department of Physics, University of Calcutta,\\
92 Acharya Prafulla Chandra Road, Kolkata - 700009, India }
\\

\end{center}
\vskip 40pt

\hspace*{\fill}
\hspace*{\fill}
\hspace*{\fill}
\hspace*{\fill}
\hspace*{\fill}
\hspace*{\fill}
\hspace*{\fill}
\hspace*{\fill}

\vskip 40pt
\hspace*{\fill}
\begin{abstract}
We present general conditions on Dirac and Majorana mass
terms under which a type-I seesaw mechanism can lead to three
exactly massless neutrinos at the tree level. We depict several
examples where the conditions are satisfied and relate some of
them to an underlying $U(1)$ symmetry. We show that higher order
corrections may generate the small observed masses and this may
be achieved even when the heavy Majorana neutrinos are at
the electroweak scale or a little higher.
\end{abstract}
   
\newpage

\section{Introduction} 

There are several good indications of  physics still
to be unveiled associated with some new high scale
above  electroweak energies. In the seesaw mechanism this idea
has been implemented for generating neutrino masses
\cite{see} which are expected to be much smaller than those for
other elementary particles. 

Here we shall consider one of the simplest and elegant versions
of the type-I seesaw mechanism where the Standard Model (SM) is
supplemented with three neutral fermion singlets ($N_{R}$). To avoid an
unattractive suppression of Yukawa couplings much
below unity, one requires the seesaw scale of the heavy Majorana
mass of $N_R$ to be of the order of $10^{8}-10^{16}$  GeV. Thus, a direct
experimental test of this idea is not possible at present.
However, there are some proposals for lowering this scale
\cite{ma,grimu1,pilaf}. One
such, in the context of the type-I seesaw mechanism, calls for a
cancellation among different contributions to the light neutrino
mass matrix.

In this work, we discuss in general terms the structure of the
neutrino mass matrix for which three exactly massless neutrinos
are possible at the tree level \cite{smi}. We show that some of
these mass matrix structures can be traced  to a $U(1)$ symmetry.
We indicate how small masses can be generated for the light
neutrinos, once this symmetry is softly broken, through higher
order effects keeping right-handed neutrinos even at the
electroweak scale.

Let us consider the model where alongside the three generations of
leptons of the SM three right-handed (SM singlet) neutrinos,
$N_{Ri} \; (i =1,2,3)$, have been added. The most general mass term for the
neutrino fields is given by

\be
  {\cal{L}}_{mass}  = {1\over 2} \left( {\bar \n_L},
{\bar{N}_R}^c \right) M
\pmatrix{\n_L^c \cr N_R} + h.c .
\label{e:lag1}
\ee
where $M$, the
$6\times 6$ neutrino mass matrix  spanning
$[ \n_e, \n_\m, \n_\tau, N_{R1}, N_{R2}, N_{R3}]$,  is 
\be
M= \pmatrix{ M_L& M_D \cr M_D^T & M_R }\;.
\label{e:massM}
\ee
It is to be noted that, in general, $M$ is not hermitian.
If there  is no Higgs triplet,
as we choose,  $M_L$ is zero in the above and there is no type-II
seesaw
contribution to the neutrino mass.
$M_D$ is related to the Yukawa couplings to scalar doublets with
vacuum expectation value. $M_R$ is the Majorana mass term and is complex symmetric
in general.  The matrix elements in $M_D$ are  much
smaller than the non-zero elements in $M_R$,
the latter in general characterizing new high scale physics.

To obtain three massless neutrinos, $M_D$ as well as $M_R$ must
be attributed with some special features.  In the following
section 2 we obtain the general conditons on $M$ which would lead
to one, two, or three exactly massless neutrinos. In section 3
we list the correlated structures of  $M_D$ and $M_R$ which
follow from the requirement of three massless neutrinos. A
discussion of how a $U(1)$ symmetry  would lead to the desired mass
matrix textures is presented in section 4. In section 5 we then
show how a small departure from the symmetry can lead to tiny
non-zero neutrino masses at the two-loop level. The higher order
nature of the correction allows the
new symmetry breaking scale to be in a relatively low range. The
testability of such a scenario is briefly outlined in section 6
before ending with our conclusions.

\section{Conditions for massless neutrinos} 

The general structure of $M_D$ and $M_R$ can be written as:
\be
M_D =  \pmatrix{x_1 & x_2 & x_3 \cr
\a_1 x_1 & \a_2 x_2 & \a_3 x_3 \cr
\beta_1 x_1 & \beta_2 x_2 & \beta_3  x_3},
\label{e:mD}
\ee
and
\be
M_R = \pmatrix{M_1 & M_4 & M_5 \cr
M_4 & M_2 & M_6 \cr
M_5 & M_6 & M_3},
\label{e:mR}
\ee
where the entries $x_i, \alpha_i, \beta_i$, and $M_i$ may be complex.
The eigenvalues of $M$ are obtained from the
characteristic equation
\be
Det\left[M^\dag M - \lambda \; diag(1,1,1,1,1,1)\right] = 0.
\label{e:char}
\ee

In this notation, we discuss next the
conditions on $M_D$ and $M_R$ for one, two, or three massless neutrinos.

\subsection{One massless neutrino} The existence of one zero eigenvalue
requires $Det(M^\dag M)$ to vanish.  This implies $A A^* =0$
where
\bea
A=Det[M_D]={\left\{\a_3 \left(\beta_2 -\beta_1 \right) + \a_2
\left( \beta_1 - \beta_3 \right) +
\a_1 \left( \beta_3 - \beta_2 \right) \right\}  }^2  x_1^2 x_2^2 x_3^2.
\label{e:onez}
\eea
This result does not depend at all on the structure of $M_R$. It
follows that the necessary condition for at least one massless
neutrino is any one of the following: (a) $\a_1=\a_2 = \a_3;\;$
(b) $\beta_1 = \beta_2 = \beta_3 ;\;$ (c) $\a_j = \a_k $ and
$\beta_j = \beta_k$ ,  $j\neq k;\;$ (d) at least one of the
$x_i$'s is zero. (a) and (b) correspond to two rows of $M_D$
being proportional, {\em i.e.,} by an appropriate redefinition of the
doublet neutrino fields one of them can be entirely decoupled.
For (c) the $j,k$ columns are proportional while in (d) one column is
vanishing\footnote{All $\alpha_i$ = 0 or all $\beta_i$ = 0 ($i
=1,2,3$) corresponds to the vanishing of a row of $M_D$.}. (d)
can be obtained from (c) by redefining the right-handed neutrino
fields; in effect, one of the three right-handed neutrinos is decoupled
for these alternatives.

\subsection{Two massless neutrinos} Demanding  the coefficient of
$\lambda$ in eq. (\ref{e:char}) to be zero along with
(\ref{e:onez}) it is possible to get two massless neutrinos.  For
brevity we do not present the expression for this coefficient
here. Instead, we list various possible solutions for two
massless neutrinos.

\hspace*{\fill}

\noindent
{\bf Solution 2.1:}
\bea
\a_1=\a_2=\a_3; \;\;  \beta_1=\beta_2=\beta_3.
\label{e:sol1}
\eea
Suitably redefining the left-handed neutrinos this results in 
two of them  being  decoupled in the mass matrix and hence massless.  \\

\noindent
{\bf Solution 2.2:}
\bea
&&x_i =0 ;   \;\; {\mbox {for some $i$ and one of the following conditions:}}  
\nonumber \\
(i)&& M_i=0  ;  \;\;\nonumber \\
(ii)&&  x_j=0 ;\; j \neq i,  \nonumber \\
(iii)&&  \a_j =\a_k ;\;\; \beta_j= \beta_k    ;\;\; i \neq j \neq k \neq i
    ;\;\;  i,j,k = 1,2,3. 
\label{e:sol2}
\eea
$(ii)$ corresponds to effectively only one coupled right-handed
neutrino and therefore only one massive light neutrino.
$(iii)$ is also the same upto a redefinition of the fields.\\  

\noindent
{\bf Solution 2.3:}
\bea
\a_1=\a_2=\a_3 ;\;\; B=0,
\label{e:sol3}
\eea
where
\bea
B =  {\left( \beta_1 - \beta_2 \right) }^2 M_3 x_1^2 x_2^2  -2 \left(
\beta_1 - \beta_2 \right) x_1 \left\{ \left( \beta_1 - \beta_3 \right)
M_6 x_1 +
\left(\beta_3 - \beta_2 \right) M_5 x_2 \right\} x_3 x_2  + \nonumber \\
\left\{ {\left( \beta_1 - \beta_3 \right)}^2  M_2 x_1^2 + 2 \left( \beta_1 -
 \beta_3
\right)  \left(  \beta_3 - \beta_2 \right) M_4 x_2 x_1+ {\left(
\beta_2 - \beta_3 \right) }^2   M_1 x_2^2 \right\} x_3^2.
\label{e:B}
\eea
Solution 2.1 is but a special case of this
one; it is listed separately for its later relevance. In general,
any $M_i$ in (\ref{e:B}) can be constrained using eq.
(\ref{e:sol3}).

\hspace*{\fill}

\noindent
{\bf Solution 2.4:}
\be              
{\mbox {This is  obtained from Solution 2.3 after replacing all} }
\;\; \a_i {\mbox { by} } \;\;\beta_i {\mbox { and vice versa.}}
\label{e:sol4}
\ee


\noindent
{\bf Solution 2.5:}
\bea
&&  \a_i =\a_j ;\;\; \beta_i= \beta_j    ;\;\; i \neq j 
    ;\;\;  i,j = 1,2,3, \;\;\; {\rm and} \nonumber \\
&& (M_R)_{ii} = (M_R)_{jj} = (M_R)_{ij} = (M_R)_{ji} = 0.
\label{e:sol5}
\eea
Thus the right-handed neutrinos of the $i,j$-type are coupled
only to the $k$-th right-handed neutrino while the coupling strengths of
the left-handed $i,j$-type neutrinos  to the right-handed
neutrinos bear the  constant ratio $x_i:x_j$.


The solutions 2.1-2.4 for two massless neutrinos presented above are
valid for general $M_R$ and as such hold for both singular or
non-singular $M_R$. The condition in 2.5 requires $Det[M_R] = 0$.

\subsection{Three massless neutrinos} To  obtain three massless
neutrinos the coefficient of $\lambda^2$ in eq. (\ref{e:char}) is
also to be zero apart from setting $\lambda$-independent  and
$\lambda^1$ terms to zero.  For every alternative for two
massless neutrinos,  {\em i.e.,} eqs.  (\ref{e:sol2}) -
(\ref{e:sol5}),  we examine what additional requirement will
yield a third massless neutrino.  Interestingly, it turns out
that, excepting for the  $2.2(ii)$ alternative on which we comment
below, in all other options solution 2.1 -- eq.  (\ref{e:sol1}) --
must be required supplemented with the following relation:
\bea
\left( M_2 M_3 -M_6^2\right) x_1^2 &+& \left( M_1 M_3-M_5^2  \right) x_2^2
+ \left( M_1 M_2 -M_4^2\right) x_3^2
+ 2 \left(M_6 M_5 - M_3 M_4\right) x_1 x_2 \nonumber \\ &+& 2
\left( M_4 M_5 - M_1 M_6 \right) x_2 x_3 + 2 \left(M_4 M_6 -
M_2 M_5 \right) x_1 x_3 = 0.
\label{e:master}
\eea
Eq. (\ref{e:master}) is the master constraint
condition on different elements of the matrix $M_R$.  For
some specific cases it restricts the different $x_i$
appearing in $M_D$;  in some alternatives one or more of the $x_i$ may
even be zero. 

In case $2.2(ii)$, where $x_i = 0$ and $x_j = 0$, one does {\em
not} require eq. (7); indeed $\alpha_{i,j}$ and $\beta_{i,j}$ cannot
even be defined. Simply satisfying  the master constraint eq.
(\ref{e:master}) is sufficient. Note that if eq.  (\ref{e:sol1})
is valid then by a suitable redefinition of the $\nu_i$ fields
$x_i = 0$ and $x_j = 0$ can be achieved.  However, the converse
is not necessary.

One may look at the condition in (\ref{e:master}) in the
following way.  If eq.  (\ref{e:sol1}) (or $x_i = 0$ and $x_j =
0$) is valid the matrix $M_D$ is of rank 1. In that case,  with
suitable unitary transformation on $M_D$ one may reduce $M$ in
(\ref{e:massM}) to a $4 \times 4$ non-zero block
\bea
M^{\prime}=\pmatrix{0 & x_1 &x_2 & x_3 \cr
x_1 & M_1 & M_4 & M_5 \cr
x_2 & M_4 & M_2 & M_6 \cr
x_3 & M_5 & M_6 & M_3 } ,
\eea
and now the massless condition of the third neutrino requires
the vanishing
of $Det [ M^{\prime^\dag} M^{\prime} ]$, {\em i.e.,} simply 
\be
Det[ M^{\prime}]=0 .
\ee
This is just eq. (\ref{e:master}).
If $M_R$ is non-singular then
another way of looking at eq. (\ref{e:master}) is to consider the 
seesaw formula
for the $3 \times 3$ light neutrino mass matrix as the power series
expansion:
\be
m_\n = - M_D M_R^{-1} M_D^T +{1 \over 2} M_D M_R^{-1} ( M_D^T
M_D^*  M_R^{-1^*} + M_R^{-1^*} M_D^\dag M_D) M_R^{-1} M_D^T +
\ldots  .
\label{e:pseries}
\ee
In that case, from (\ref{e:pseries}) one can see the massless
condition for all three neutrinos to all orders will correspond
to the vanishing of the leading order term \cite{grimu}, {\em
i.e.,}
\be
 M_D M_R^{-1} M_D^T=0 .
\label{e:lead} 
\ee
If we use eq. (\ref{e:sol1}) on the left-hand side of this equation then
we get a $3 \times 3$ matrix each element of which has a common
factor.  This common factor is nothing but the left-hand-side of
eq. (\ref{e:master}). Thus,   (\ref{e:lead}) together with
(\ref{e:sol1}) implies eq. (\ref{e:master}).\\

\section{Correlated Dirac and Majorana sectors}

Next, based on eqs.  (\ref{e:sol1}) and (\ref{e:master})  we
discuss various possible structures of $M_R$ and $M_D$ that
together result in three massless neutrinos. We consider cases
where one or more of the $x_i$ are non-vanishing. We ask the
question what can one say about $M_R$ for three massless
neutrinos for a chosen form of $M_D$ {\em irrespective of the
specific non-zero values of $x_i$}.  Determinant of $M_R$ is
permitted to be zero or non-zero.  We discuss them separately.

It is noteworthy that if any two of the $x_i$ are nonzero (and
certainly if all three are non-zero) then for arbitrary values of
these $x_i$ there is no solution satisfying eq.  (\ref{e:master})
unless $Det[M_R] = 0$.   Such examples are taken up after
considering the only option when $Det[M_R]$ can be non-vanishing.
\\

\noindent
\subsection{The ${\bf Det[M_R] \neq 0}$ case} There is only one
class of possibilities in this category.

\vskip 10pt

{\bf (i) Only one ${\bf x_i}$ nonzero:} Considering $x_1=x_2=0$
in $M_D$ and $x_3$ is arbitrary, (\ref{e:master}) implies the
following forms for $M_D$ and $M_R$:
\bea
M_D =  \pmatrix{0 & 0 & x_3 \cr
0 & 0 & \a x_3  \cr
0 & 0 & \beta x_3 },
 \;\; M_R = \pmatrix{M_1 & \pm \sqrt{M_1 M_2} & M_5\cr
\pm \sqrt{M_1 M_2 } & M_2 & M_6 \cr
M_5 & M_6 & M_3  },  
\label{e:cond1}
\eea
provided that
\be
M_6 \sqrt{M_1} \neq \pm M_5 \sqrt{M_2}\;\;.
\label{e:detno}
\ee 
Above, the $\pm$ sign is chosen matching the $\pm$ sign in
(\ref{e:cond1}).  This inequality condition is required only to
satisfy $Det[M_R] \neq 0$. One may choose  any of $M_1   , \;
M_2, \; M_5, \; M_6$ equal to zero keeping in mind eq.
(\ref{e:detno}).  There is no condition on $M_3$ and it may take
zero or non-zero values.  The two different signs
in the off-diagonal elements signify the possibility of
different CP phases of heavy Majorana neutrinos satisfying the
three massless neutrino condition.

One may consider  $x_1=x_3=0$ or 
$x_2=x_3=0$  in (\ref{e:master}) to find other possible forms of
$M_R$. They follow from the previous example through
suitable permutations. For $x_2=x_3=0$ in $M_D$ and
$x_1$ arbitrary the possible form
of $M_R$ is  obtained by replacing $M_6$ by
$\sqrt{M_2 M_3}$ and satisfying the inequality $M_4 \sqrt{M_3}
\neq \pm M_5 \sqrt{M_2}$; one may consider some of  $M_2,  \;
M_3, \; M_4, \; M_5$ equal to zero. There is no condition on
$M_1$. One may note that the cancellation structure of $M_R$
given in eq. (26) in ref. \cite{smi} can be obtained in this case
if one sets $M_2=M_5=0$.  Similar solutions exist for 
$x_1=x_3=0$ in $M_D$.

It is important to bear in mind that in all the above cases eq.
(\ref{e:sol1}) cannot be imposed.\\

\noindent
\subsection{The ${\bf Det[M_R]=0}$ case}
The conventional see-saw formula breaks down when $Det[M_R]=0$.
It needs to be stressed, however, that the formula is but an
approximation. Here we are using the diagonalisation of the full
$(6 \times 6)$ neutrino mass matrix, without taking recourse to
the see-saw formula, to arrive at the mass
eigenvalues and therefore $Det[M_R]=0$ causes no difficulty. 

{\bf (i) Only one ${\bf x_i}$ nonzero:} In the cases where two of
the $x_i$ are zero (discussed above) if we replace the
inequalities, {\em e.g.,} in (\ref{e:detno}), by equality sign
then they correspond to $Det[M_R]=0$. This would imply
correlations within the elements of $M_R$ beyond what is
necessary and sufficient for three massless neutrinos.

{\bf (ii) Two ${\bf x_i}$ nonzero:}  Next, let us consider cases
when only one of the $x_i$ is zero.  As for example, choosing
$x_1\neq 0,\;x_2\neq 0,\;x_3=0$ in (\ref{e:master}),
three massless neutrinos for arbitrary choices of $x_{1,2}$
requires that only the (12) block of $M_R$ be non-zero:
\bea
M_D =  \pmatrix{x_1 & x_2 & 0 \cr
\a x_1 & \a x_2 & 0 \cr
\beta x_1 & \beta x_2 & 0},
\;\; M_R = \pmatrix{M_1& M_4&0\cr
M_4&M_2&0\cr
0&0&0}.
\label{e:0det12}
\eea
Similarly, for only $x_2=0$ the (13) block  and for
only  $x_1=0$  the (23) block of $M_R$ is non-zero.

{\bf (iii) All ${\bf x_i}$ nonzero:} Finally, when in addition to
the requirement of eq.  (\ref{e:sol1}) {\em all} $x_i$  in $M_D$
are non-zero then the solution for $M_R$ is given by:
\bea
M_D =  \pmatrix{x_1 & x_2 & x_3 \cr
\a x_1 & \a x_2 & \a x_3 \cr
\beta x_1 & \beta x_2 & \beta  x_3},
\;\; M_R= \pmatrix{M_1 & \pm \sqrt{M_1 M_2} & \pm\sqrt{M_1 M_3}\cr
\pm \sqrt{M_1 M_2}& M_2 &\pm \sqrt{M_2 M_3}\cr
\pm \sqrt{M_1 M_3}&\pm \sqrt{M_2 M_3}&M_3}   \; .
\label{e:detyes}
\eea
Above, alternate sign choices reflect the possibility of different
CP phases of heavy Majorana neutrinos. The democratic form of
$M_R$ emerges as a special case when $M_1=M_2=M_3$ with positive
sign in the off-diagonal elements in (\ref{e:detyes}).

In summary, for the most general possible form of $M_R$ with
$Det[M_R] \neq 0$ the type of solution is given by
(\ref{e:cond1}) where $M_D$ is highly constrained with two $x_i$
being zero. On the other hand, if we opt for the most general
form of $M_D$ consistent with solution 2.1 then the form for
$M_R$ is given by (\ref{e:detyes}) in which all off-diagonal
elements are fixed in terms of the diagonal entries and $Det[M_R] = 0$.
(\ref{e:0det12}) is an intermediate situation between these
extremes. \\

\section{A $U(1)$ symmetry} 

So far we have outlined the manner in which
$M_D$ of eq. (\ref{e:mD}), already restricted through eq.
(\ref{e:sol1}),  and $M_R$ of (\ref{e:mR}) are further
constrained through eq.  (\ref{e:master}) in order to obtain
three massless neutrinos. The natural next question is whether
there is any symmetry at the root of the cancellation of
different contributions to the light neutrino mass matrix, thus
rendering  three neutrinos massless.  Below we exhibit a $U(1)$ symmetry
which could imply the textures of $M_D$ and $M_R$ in eqs.
(\ref{e:cond1}), (\ref{e:0det12}) and (\ref{e:detyes}). In the next section,
we
further show that in this $U(1)$ model when the symmetry is
softly broken, the loop level corrections generate small
neutrino masses.

We consider a
$U(1)$ symmetry  of the basic Lagrangian \cite{branco}.
Following ref. \cite{ma} a new scalar doublet $\chi=
\pmatrix{\chi^+ \cr \chi^0}$  is introduced with $U(1)$ quantum
number +1. The Standard Model (SM) doublet Higgs, $\phi$, has
$U(1)$ charge 0. All SM quarks and leptons are assigned the same
$U(1)$ charge (=1) so that they receive their masses through the
Yukawa coupling to the $\phi$, while a coupling to $\chi$ is
forbidden for them. Assigning zero $U(1)$ charge to the three neutral
fermion singlets  $N_{iR}$, such a symmetry has been used in
\cite{ma} to forbid the neutrino Dirac masses from arising from
$\phi$ and only $\chi$  can contribute here.  Arranging $\bra
\chi \ket \ll \bra \phi \ket$ the smallness of the neutrino Dirac
mass is ensured. As we discuss in the following, in this work too
we use the scalar $\chi$ to generate small neutrino Dirac masses.
Further, we choose the $U(1)$ charges of the $N_{iR}$
appropriately to reproduce the desired textures of $M_D$ and
$M_R$. Thus the $U(1)$ symmetry serves a dual role.

We consider the following $U(1)$ transformations:
\bea 
L \rightarrow e^{i \gamma n_L } L ; \;\; l_R \rightarrow e^{i
\gamma n_R} l_R;\;\;
N_R \rightarrow e^{i \gamma n_{\n}} N_R ,
\label{e:sym1}
\eea 
where $\gamma$  is real and $n_L$, $n_R$ and $n_{\nu}$ are
hermitian matrices acting on flavor space. $L_j$ are the
left-handed lepton doublets, $l_{Rj}$ the right-handed charged
lepton singlets, and $N_{Rj}$ the right-handed neutrino fields.
Except these and the scalars ($\phi$, $\chi$) no other fields transform under
this $U(1)$. 
In the basis where the charged lepton mass
matrix is diagonal one can take
$n_L =n_R = diag(n_1,n_2,n_3)$  where $n_{1,2,3}$ are $U(1)$
charges of $e,\;\mu \;$, and $\tau$ respectively.

The lepton sector masses arise from the Lagrangian:
\bea
 {\cal L_{Y}} =  -  Y_{ij} \bar{L}_i \phi l_{Rj} - Y^{\nu}_{ij}
{\bar L}_i \tilde{\chi}    N_{Rj} - {1 \over 2} {\bar N^c_{Ri}}
{M_R}_{ij} {N_{Rj}},\;\; 
+ \;\; h.c.
\label{e:lag2}
\eea
where  $\tilde{\chi} = i\sigma_2 \chi^*$. The assignments of
$n_L$, $n_R$ and $n_{\nu}$ determine the non-vanishing elements
of $Y_{ij}$, $Y^{\nu}_{ij}$ and ${M_R}_{ij}$.  With only $U(1)$
symmetry one cannot derive relationships among the non-zero Yukawa
couplings $Y^{\nu}_{ij}$. We now take up the different cases in turn.\\

{\bf (i) Only one ${\bf x_i}$ nonzero:} Here the mass matrices,
$M_D$ and $M_R$ will be as in (\ref{e:cond1}) when $x_1 = x_2 =
0$.  Choosing, for example,
\be (i) \;
n_{\nu}=diag(-2,0,2)   ;\;\; (ii)  \; n_{\nu}  =  diag(0,-2,2) ;\;\;
(iii) \;  n_{\nu}= diag(-2,-2,2) ,
\label{e:sym2}
\ee 
one can respectively obtain the mass matrices
\be 
(i)\;  M_R=
\pmatrix{0 & 0&M_5 \cr 0 & M_2 & 0 \cr M_5 & 0 &0 };\;(ii) 
\; M_R=\pmatrix{ M_1&0&0\cr 0&0&M_6 \cr 0& M_6 &0 };\;   
(iii)\;  M_R=\pmatrix{0&0&M_5 \cr 0&0& M_6 \cr M_5 & M_6 & 0
},
\label{e:mforms}
\ee 
all of which are in the class of (\ref{e:cond1}). Of these,
$Det[M_R]$ is non-vanishing for the first two while it is zero for
the third. (\ref{e:sym2}) along with the choice: 
\be 
n_L =n_R = diag(1,1,1)
\label{e:sym3}
\ee 
ensures that  $\bra \chi \ket$ reproduces the $M_D$ in
(\ref{e:cond1}) and that $\bra \phi \ket$ does not contribute.
These three basic textures are the most general possibilities
apart from trivial permutation of various Majorana neutrino
fields.

As noted after eq.  (\ref{e:cond1}), there are solutions
where rather than ($x_1, x_2$)  other pairs vanish.  It is
simple to change the $U(1)$ quantum number assignments in
(\ref{e:sym2}) to achieve these forms of $M_D$ and $M_R$.  
One set, {$x_2 \neq 0$}, gives a previously not discussed
structure for three massless neutrinos while another, {$x_1
\neq 0$}, reproduces the model obtained earlier in ref.
\cite{smi}.

{\bf (ii) Two ${\bf x_i}$ nonzero:}  Let us now turn to the $M_D$
and $M_R$ in (\ref{e:0det12}). Here, two $x_i$ in $M_D$ are
non-zero. One gets these $M_R$ and $M_D$ 
with, for example, the assignments $n_{\nu}=diag(0,0,2)$ and
$n_L=n_R=diag(-1,-1,-1)$ to get {$x_3 = 0$}.

{\bf (iii) All ${\bf x_i}$ nonzero:} The form of $M_R$ in
(\ref{e:detyes}) requires the choice  $n_{\nu}=diag(0,0,0)$. This
along with $n_L=n_R=diag(-1,-1,-1)$ reproduces the desired
texture of $M_D$. Of course, the relationships between the matrix
elements cannot be obtained through the $U(1)$ symmetry.  

At this stage it is worth noting that the forms of $M_D$ and
$M_R$ in eq. (\ref{e:detyes}) lead to three massless neutrinos
for {\em arbitrary} values of $x_{1,2,3}$ and $M_{1,2,3}$. A
particular choice, $x_3 = 0$ and $M_1 = M_2 = 0$, i.e., 
\bea
M_D =  \pmatrix{x_1 & x_2 & 0 \cr
\a x_1 & \a x_2 & 0 \cr
\beta x_1 & \beta x_2 & 0},
\;\; M_R = \pmatrix{0 & 0 & 0 \cr
0 & 0 & 0 \cr
0 & 0 & M_3} \; ,
\label{e:detyes2}
\eea
can be
accomplished by the $U(1)$ symmetry by choosing $n_L = n_R =
diag(1,1,1)$ and $n_\nu = diag(2,2,0)$. Needless to say, other
similar special cases of eq. (\ref{e:detyes}) with $x_2 = 0$, $M_3 =
M_1 = 0$ and  $x_1 = 0$, $M_2 = M_3 = 0$ can also be derived from
the $U(1)$ symmetry.  \\

\section{Non-zero neutrino masses} 

Present experimental  data,
{\em e.g.,} direct mass measurements, neutrino oscillations,
etc.,  indicate that light neutrinos  have a mass below about 0.1
eV. To generate such tiny masses it is required to break the
above noted $U(1)$ symmetry  of ${\cal L_{Y}}$
-- eq. (\ref{e:lag2}).  Particularly, one may include a soft
symmetry breaking  term
\cite{ma}:
\bea
\mu_1^2 \left( \phi^\dag \chi + \chi^\dag \phi \right).
\label{e:Lviol}
\eea
This would result in  charged physical Higgs bosons given by
\bea
h^{\pm}  = {v \chi^{\pm} - u \phi^{\pm}   \over \sqrt{v^2 + u^2}},
\eea
where $v = \bra \phi^0  \ket $ and $ u = \bra \chi^0 \ket$ and
the $U(1)$ conserving  terms in the scalar potential $V(\phi,
\chi)$ can be chosen\footnote{See, for example, eq.  (11) in ref.
\cite{ma}.} so as to ensure $u \ll v$. The $U(1)$ charge
assignments require the quark and charged lepton masses to arise
through $v$ while the smaller neutrino masses originate from $u$.
The $U(1)$ violating soft interaction (\ref{e:Lviol}) will induce
contributions to $M_D$ and $M_R$ from higher order corrections
which may result in deviations from the textures responsible for
three massless neutrinos and give rise to small neutrino masses.
Here, we pick two specific textures of $M_D$ and $M_R$ given in
eqs. (\ref{e:detyes})  and (\ref{e:detyes2}) which result in
three massless neutrinos. We consider one- and two-loop
corrections to these textures and show how light neutrino masses
and splittings are generated. The loop corrections are calculated
in the {\em weak} basis, i.e., to $M_D, M_L$, and $M_R$, and the
consequences that follow are explored.

\begin{figure}[bht]
\begin{center}
\hskip -0.0cm
\vskip -0.8cm
\psfig{figure=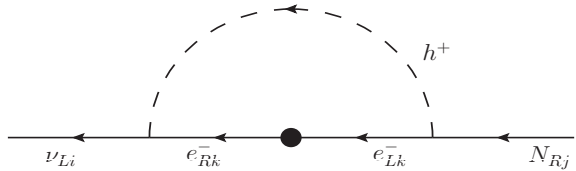,width=14.0cm,height=18.0cm,angle=0}
\vskip -12.5cm
\caption{ \sf \small
One-loop Feynman diagram contributing to $M_D$.}
\label{f:1loop}
\end{center}
\end{figure}

\noindent
\subsection{One-loop effects} At one-loop level a typical
contribution is the diagram in Fig. \ref{f:1loop} which
contributes to $M_D$.  This is given by:
\bea
\Delta {M_D}_{ij}^{(1)} \approx \sum_{k}
{Y_{{ik}} \; Y^{\nu}_{kj}  \over {32 \; \pi^2  }} {m_{e_k} \over
m_{h^+}^2}\;  \sin 2 \kappa \; \left[ m^2_{e_k} \;
{\overline{\ln}} (m^2_{e_k}) - m^2_{h^+} \;{\overline{\ln}}
(m^2_{h^+})
\right]
\; ,
\label{e:1loop}
\eea
where  $\overline{\ln} (x) =  \ln (x/Q^2)$,  $Q$ being the
renormalization scale, and
\be
\sin \kappa = u/\sqrt{v^2 + u^2 }.
\ee
In spite of breaking the $U(1)$ symmetry softly, however, this
one-loop effect does not change the texture of the matrix
$M_D$ as we note below.

Recall from  eq.  (\ref{e:lag2}) that at  the tree level,
${M_D}_{ij} = {Y_{ij}^{\nu}} \;u$.  To satisfy eq.
(\ref{e:sol1}), which is a key requirement for  massless neutrinos, one
must have ${M_D}_{2j}/{M_D}_{1j} = Y_{2j}^{\nu}/Y_{1j}^{\nu}$ and
${M_D}_{3j}/{M_D}_{1j} = Y_{3j}^{\nu}/Y_{1j}^{\nu}$ to be $j$
independent\footnote{This can be represented as $Y_{ij}^{\nu} =
f_i Y_{1j}^{\nu}$ where $f_{(1,2,3)} \equiv (1, \alpha,
\beta)$.}.  This relationship is preserved after inclusion of the
one-loop corrections.

To see this, note that the contribution from eq.
(\ref{e:1loop}) is $\Delta {M_D}_{ij}^{(1)} = \Sigma_k X_{ik}f_k
Y_{1j}^{\nu}$, where $X_{ik}$ encapsulates the entire factor
multiplying $Y_{kj}^{\nu}$. Thus, $\left( {M_D}_{ij} + \Delta
{M_D}_{ij}^{(1)} \right) / \left( {M_D}_{kj} + \Delta
{M_D}_{kj}^{(1)} \right)$ remains independent of $j$.

There is a similar correction to $M_D$ from $h^0$ exchange.
In this, or a similar one due to $W^\pm$ exchange,
there is an overall factor of $M_D$. Thus this additional piece
will also leave the texture of $M_D$ unchanged and is not pursued
any further.

Likewise, there will be one-loop contributions to $M_L$ through
$h^0$ or $W^\pm$ exchange. These are proportional to $M_L$.
Since we have chosen $M_L = 0$ at the tree level the loop
corrections  also vanish.

There are diagrams similar to Fig. \ref{f:1loop} involving
$W^{\pm}$ exchange contributing to $M_R$ \cite{petcov}. These
corrections involve the couplings between $\nu_L$ and $N_R$ 
arising from the second term of ${\cal
L_{Y}}$ in (\ref{e:lag2}) and are proportional to $\bra \chi
\ket$. In general, they are expected to be very small. A larger
contribution may arise from the one loop diagram similar to Fig.
\ref{f:1loop} with $h^+$ exchange with mass insertion  on the
external leg. The  one-loop contribution to $M_R$  due to this
is given by (assuming $m_{h^+}^2 > M_{Rij}^2 $)
\bea
\Delta {M_R}_{ij}^{(1)} \approx \sum_{k,m}
C_1 \; {Y_{kj}^\nu Y_{mk}^{\nu \dag} \cos^2 \kappa  \over {16 \; \pi^2
 m_{h^+}^2 }}
 \; \left[
m_{h^+}^2 + M_{Rij}^2 + m^2_{e_k} {m_{h^+}^2 \over M_{Rij}^2  }
\; {\overline{\ln}} (m^2_{e_k}) - m^2_{h^+}
\;{\overline{\ln}} (m^2_{h^+})
\right]  M_{Rim} 
\; 
\label{e:mr1loop}
\eea
where
\bea
 C_1 & \approx &  1  \;\; \mbox{for} \; \; M_{Rn}  << M_{Rij}  \nonumber \\
& \approx &   \frac{{M_{Rij}}}{M_{Rn}}  \nonumber  \; \; \mbox{otherwise}
\eea
and $M_{Rn}$ is a typical eigenvalue of $M_R$ .

Let us now turn to the effect of these corrections on the light
neutrino masses. Consider first the $M_D$ and $M_R$ given in
eq. (\ref{e:detyes2}). Incorporating the one-loop corrections
these become
\bea
M_D =  \pmatrix{x'_1 & x'_2 & 0 \cr
\alpha ' x'_1 & \alpha ' x'_2 & 0 \cr
\beta ' x'_1 & \beta ' x'_2 & 0},
\;\; M_R = \pmatrix{M'_1 & M'_4 & 0 \cr
M'_4 & M'_2 & 0 \cr
0 & 0 & M'_3} \; ,
\label{e:detyes2a}
\eea
where the primes indicate the one-loop corrected values. Notice
that the form of (\ref{e:detyes2a}) satisfies the conditions in
2.2(iii) and thus will lead to {\em two} massless neutrinos.
Therefore, one must check whether two-loop effects can remove the
remaining degeneracy.

Similarly, for the $M_D$ and $M_R$ given in
eq. (\ref{e:detyes}) inclusion of the one-loop contributions
results in
\bea
M_D =  \pmatrix{x'_1 & x'_2 & x'_3 \cr
\alpha ' x'_1 & \alpha ' x'_2 & \alpha ' x'_3 \cr
\beta ' x'_1 & \beta ' x'_2 & \beta ' x'_3},
\;\; M_R= \pmatrix{M'_1 & M'_4 & M'_5\cr
M'_4& M'_2 &M'_6\cr
M'_5& M'_6&M'_3}   \; .
\label{e:detyesa}
\eea
As in the previous example, the effect of these corrections is to
make one of the neutrinos massive. $M_D$ in (\ref{e:detyesa})
satisfies the condition 2.1 for {\em two} massless neutrinos.

The result that {\em one} of the neutrinos acquires mass through
the one-loop corrections is similar to that of \cite{pilaf} where
a model with only SM interactions and  right-handed singlet
neutrinos is examined.  In order to remove the remaining neutrino
mass degeneracy, as required by the atmospheric and solar
observations,  we turn to two-loop effects now. In both examples
above, the degeneracy  is due to the textural property of $M_D$.
So, it is enough to focus on the two-loop effects on this matrix.

\subsection{Two-loop effects} 
\begin{figure}[htb]
\begin{center}
\hskip -1.0cm                             
\vskip -1.2cm              
\psfig{figure=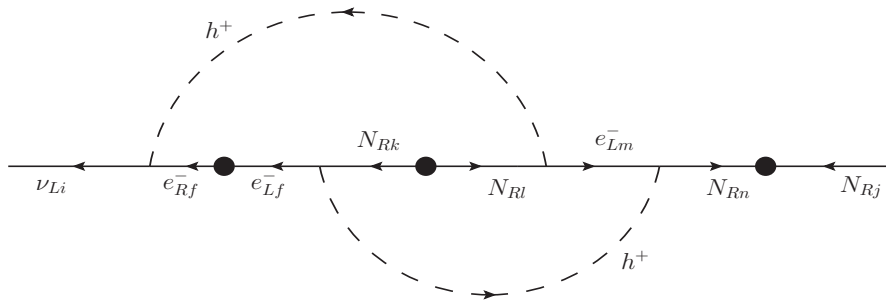,width=14.0cm,height=18.0cm,angle=0}
\vskip -11.5cm
\caption{ \sf \small
Two-loop Feynman diagram contributing to $M_D$.
}
\label{f:2loop}
\end{center}
\end{figure}

We find that at the two-loop level
there are diagrams which yield contributions which deviate from
eq.  (\ref{e:sol1}). Consider, for example,  the two-loop Feynman
diagram in Fig. \ref{f:2loop} (in which $N_k$ and $N_l$ are the $k$-th and
$l$-th flavour eigenstates).
The contribution to $M_D$ from such diagrams  is estimated to be:
\bea
&&\hspace{-0.6cm} \Delta {M_D}_{ij}^{(2)} \approx  {C_2 \over {(4
\pi)}^4}\sum_{f,k,l,m,n}\;
Y_{{if}} \;
Y^{\nu }_{fk} \; Y^{\nu }_{ml} \;
{Y^{\nu * }_{mn}}\; \sin \; \kappa \;
\cos^3\kappa \; m_{e_f} {M_R}^*_{kl} {M_R}_{nj} \
\nonumber \\ 
&&\hspace{1.2cm} \; 
Re\left[ I_1(m_{h^+}^2,m_{e_m}^2,m_{e_f}^2,m_{h^+}^2,M_{Rn}^2)
\right] ,
\eea
where
\bea
 C_2 & \approx &  1  \;\; \mbox{for} \; \; M_{Rn}  << M_{Dij}  \nonumber \\
& \approx &   \frac{{M_{Dij}}}{M_{Rn}}  \nonumber  \; \; \mbox{otherwise}
\eea
and $M_{Rn}$ is a typical eigenvalue of $M_R$ . In the above expression, 

\bea
&&I_1(m_{h^+}^2,m_{e_m}^2,m_{e_f}^2,m_{h^+}^2,M_{Rn}^2)=  \nonumber \\
&& A^2 \;\int d^d k \int d^dq {1 \over {(k^2+
m_{h^+}^2) (q^2+ m_{e_m}^2) ((k-p)^2+m_{e_f}^2)((q-p)^2+m_{h^+}^2)((k-q)^2+
M_{Rn}^2)}}\; ,
\nonumber \\
\eea 
and $A={(2 \pi \mu)}^{2 \e}/\pi^2$ is the loop factor and
integrals are regularized by dimensional reduction to $d = 4 - 2
\e$ dimensions.

To estimate \cite{mart} the integral $I_1$ we shall ignore the
external invariant momentum $-p^2$ (as ${M_D}_{ij}$ is small) and
charged lepton masses with respect to the scalar, $h^\pm$,  and
right-handed neutrino masses.  Then the integral $I_1$ in (34)
can be approximated (in ${\overline{MS}}$ scheme) for $m^2_{h^+}
> M_{Rn}^2$ as
\bea
Re[ I_1] \approx {1 \over x^2}  \left[ 2 (x-y)  \left\{ Li_2
(y/x) - {\overline{\ln}} (x-y)\; \ln(x/y)\right\} +(x-3 \; y/4)
{({\overline{\ln}} (x) )}^2
    \nonumber \right. \\   \left.   + (2 x + y)  {\overline{\ln}}
(x) \; {\overline{\ln}} (y)
 +  y \; \zeta (2)
+ (y/2) \; {({\overline{\ln}} (y) ) }^2 \right] ,   
\eea
where $x={m^2_{h^+}}$, $y= M_{Rn}^2$.
For $m^2_{h^+} < M_{Rn}^2$ it can be approximated as
\bea
Re[ I_1] \approx {1 \over x^2}  \left[ 2 (y-x)  \left\{ Li_2 (x/y)
- {\overline{\ln}} (y-x)\; \ln(y/x)\right\}
+(3 y/2 - \; x) \times
    \nonumber \right. \\   \left.
\left\{ {({\overline{\ln}} (y) )}^2 +2 \zeta (2) \right\}
    - y {\overline{\ln}} (x) \; {\overline{\ln}} (y)
- (y/2-x) \; {({\overline{\ln}} (x) ) }^2 \right] \; .   
\eea
In the asymptotic limit \cite{babu}
\bea
I_1&\approx& {\pi^2 \over 3 x}
 \; \; \; \mbox{for} \; m^2_{h^+} \gg M_{Rn}^2\; ,  \nonumber \\
   &\approx& {  (\ln^2(y/x) + \pi^2/3 -1)  \over   y    }
\; \; \; \mbox{for} \; m^2_{h^+} \ll M_{Rn}^2 .
\eea
One can see that unlike the one-loop result,  the two-loop
contribution can change the texture of $M_D$ as the
$j$-dependence emerges from ${M_R}_{nj}$ as well as from
$Y_{nj}^{\nu}$. So the relationships $\a_1 = \a_2 = \a_3$ and $\b_1 =
\b_2 = \b_3$ will cease to apply. The novelty  of neutrino mass generated
 in this manner
is that there is a seesaw
mechanism operative and then there is suppression\footnote{Note
that such a suppression occurs at the tree level in ref \cite{ma}
but there is further reduction here as the contribution leading
to non-zero light neutrino masses arises not
from tree level Yukawa couplings nor from one-loop but from
two-loop diagrams.} due to $\sin \kappa$ which is approximately
proportional to  $ \bra \chi^0 \ket \over {\bra \phi^0 \ket}$. 

We now proceed to examine the effect of these contributions on
neutrino masses. Consider first the $M_D$ given in eq.
(\ref{e:detyes2}). With two-loop corrections it is
\bea
M_D =  \pmatrix{x''_1 & x''_2 & 0 \cr
\alpha_1 x''_1 & \alpha_2 x''_2 & 0 \cr
\beta_1 x''_1 & \beta_2 x''_2 & 0},
\label{e:detyes2b}
\eea
where now $\alpha_1 \neq \alpha_2, \;\; \beta_1 \neq \beta_2$,
and the double primes indicate two-loop corrected values. This
form of $M_D$ results in {\em one} massless neutrino ($x_3 = 0$).
We have used the $M_D$ of eq. (\ref{e:detyes2b}) and $M_R$ of
eq. (\ref{e:detyes2a}) to obtain the light neutrino masses.
Though two light neutrinos are indeed massive, we find that it is
not possible to reproduce the observed mass splittings in this case.

On the other hand, the second example we have been considering,
namely  $M_D$ and $M_R$ in (\ref{e:detyes}), can produce a mass
spectrum in line with observations.  Including one- and
two-loop corrections, one has
\bea
M_D =  \pmatrix{x''_1 & x''_2 & x''_3 \cr
\alpha_1 x''_1 & \alpha_2 x''_2 & \alpha_3  x''_3 \cr
\beta_1 x''_1 & \beta_2 x''_2 & \beta_3 x''_3},
\;\; M_R= \pmatrix{M''_1 & M''_4 & M''_5\cr
M''_4& M''_2 &M''_6\cr
M''_5& M''_6&M''_3}   \; .
\label{e:detyesb}
\eea
with $\alpha_1 \neq \alpha_2\neq \alpha_3 \neq \alpha_1, \;\;
\beta_1 \neq \beta_2 \neq \beta_3 \neq \beta_1$.

We find that the $M_D$ and $M_R$ in eq. (\ref{e:detyesb}) do
yield the correct neutrino mass spectrum when the various
parameters are assigned appropriate values. As a typical example,
the parameters may be chosen as given in the following.

The $U(1)$ symmetry is broken in the scalar potential.  As
discussed in \cite{ma}, keeping  the strength of the soft
breaking term in eq. (\ref{e:Lviol}) $\mid \mu_1^2 \mid ~\approx$
10 GeV$^2$, the remaining terms in the scalar potential can be
chosen such that $u = \bra \chi^0 \ket \sim$ 1 MeV.  Recalling
that  $v = \bra \phi^0 \ket \sim$ 174 GeV, this implies $\sin
\kappa \sim 5.7 \times 10^{-6}$.  We consider $m_h^+ \sim$ 250
GeV and $m_\chi$ = 200 GeV. We keep the Yukawa couplings $Y_{ij}^\nu$ to be
about ${\cal O} (10^{-3})$ and $Y_{ij}$ is lesser than about 0.1.
The product of these couplings with
 $u$ sets the scale for the
tree-level entries of $M_D$.


\begin{table}
\begin{center}
\begin{tabular}{|c|c|c|c|}
\hline
Parameter&\multicolumn{3}{c|}{$i$}\\\cline{2-4}
& 1 & 2 & 3  \\ \hline
$x_i$ (in eV) & 158 & 564 &  640 \\ \hline
$\alpha_i$&  $ 10^1$ & $ 10^1  + 5.12 \times 10^{-4}$ &
 $ 10^1 + 7.68 \times 10^{-4}$
\\ \hline
$\beta_i$&   $1.2 \times 10^{1}$ & $ 1.2 \times 10^{1} + 3.84 \times 10^{-4}$
& $1.2 \times 10^{1} + 2.56
\times 10^{-4}$ \\ \hline
\end{tabular}
\caption{The parameter choices for $M_D$. The tree level term and
the higher order corrections are indicated separately.}
\label{t:md}
\end{center}
\end{table}

Including the higher order contributions the parameters defining
$M_D$  are shown in Table \ref{t:md}. The matrix $M_R$, 
including loop corrections, is given in eq. (\ref{e:detyesc}).
\bea
M_R =  \pmatrix{ 9 \times 10^{10} + 1700 &  9.48683 \times 10^{10} + 1000 & 
 1.08167 \times 10^{10} + 51200 \cr
9.48683 \times 10^{10} + 1000 &  1 \times 10^{11}  + 2500 &
1.14018 \times 10^{11} + 1500 \cr
1.08167 \times
10^{11} + 51200 &  1.14018 \times 10^{11} + 1500 &  1.3 \times 10^{11}
+ 1000  } ~{\rm eV}.
\label{e:detyesc}
\eea
$M_R$ in (\ref{e:detyesc}) corresponds to $M_R$ in (\ref{e:detyes})
with $M_1 = 90 $   GeV, $ M_2 = 100 $ GeV and $M_3 = 130 $ GeV.
One may note here that in our model
due to the loop
suppression as well as the $\sin\kappa$ factor in the light neutrino
mass, the mass scale for heavy right handed neutrinos - the seesaw scale -
could be as low as the standard electrowek scale. This is the main focus of
our work.

Here the one loop
corrections to $M_R$ make $Det[M_R]$ non-zero
and large and as such
one may use type-I seesaw formula
to obtain the light neutrino masses.
With these choices the light neutrino masses are about 0.045
eV,  0.0092 eV and 0.0 eV reproducing the two mass squared
differences of about $8 \times 10^{-5}$ eV$^2$ and $
2 \times 10^{-3}$ eV$^2$ respectively in the normal hierarchical
neutrino mass pattern.
The one loop correction to massless texture of $M_R$ and two loop
corrections to massless texture of $M_D$ essentially set the
two different scales
of mass squared differences of light neutrinos.
However, these specific choices of parameters
do not reproduce appropriate mixing.
We have not made an exhaustive survey and expect that other
parameter choices may lead to even more acceptable solutions.\\

\section{Testability} 

The above estimations indicate that in
these models the charged higgs, $h^\pm$, and the right-handed
neutrinos, $N_{Rj}$, could well be within the range of the LHC.
Depending on the ordering of $m_{h^{\pm}}$ and  $M_{Rn}$ the
signals would be different. If $m_{h^\pm} > M_{Rn}$ then on
Drell-Yan pair production of $h^\pm$ one may expect the
observable decay chain $h^{\pm}
\rightarrow l_i^{\pm} N_{Rj} $ followed by $N_{Rj} \rightarrow
l_k^{\pm} W^{\mp}$. The decay of $h^\pm$ may not be much
suppressed.  If the $N_{Rj}$ is
long-lived due to small mixing  the right-handed neutrinos may
well decay outside the detector; so just a pair of oppositely
charged leptons with missing energy will be observed.  Otherwise,
the Majorana nature of $N_{Rj}$ can lead, in addition to a
$W^{\pm}$ pair, to four leptons of which three may be of same
sign. From such signatures at the collider the strength of the coupling
$Y_{ij}^\nu$, which plays significant role in determining the
neutrino mass, and also $Y_{ij}$ may be estimated.  On the other
hand, if $m_{h^\pm} < M_{Rn}$ then one must consider $N_{Rj}
\rightarrow h^{\pm} l_i^{\mp}$, if right-handed neutrinos are
produced {\em via} their small mixing with $\nu_L$
\cite{he}.  The more important signal in this case will be
through Drell-Yan pair production of $h^{\pm}$ which will lead
to two charged tracks with matching $p_T$ since the decay $h^\pm
\rightarrow l_i^\pm \nu_j$ is suppressed by ${\sin^2 \kappa}$.
The above indicates that there is
a possibility to cross-check the neutrino mass parameters from
neutrino oscillation experiments and from collider signatures.\\

\section{Summary} The main emphasis of our paper is twofold:
(a) We identify the Dirac and right-handed Majorana neutrino mass
textures which lead to one, two, or three massless neutrinos. (b)
We demonstrate that one may not need  a new high seesaw scale for
small neutrino mass.  Right-handed neutrino fields  even at the
electroweak scale and a scalar doublet with a  $U(1)$
symmetry could accomplish this naturally. When the $U(1)$
symmetry is broken, one of the massless neutrinos acquires a mass
at the one-loop level while the remaining two  become massive
when two-loop
contributions are included. Since the two small mass splittings
arise at different orders in perturbation theory their relative
sizes can be reproduced. These ideas admit
exploration at the LHC.

\hspace*{\fill}

\hspace*{\fill}

\noindent
{\large {\bf {Acknowledgments}}}\\

R.A. acknowledges the hospitality of Harish-Chandra Research
Institute under the XI-th Plan project on `Neutrino Physics'
while this work was done. A.R. also acknowledges partial support
from this project as well as RECAPP. Both R.A. and A.R. like to
thank organisers of the WHEPP-XI workshop held at PRL,
Ahmedabad, India where some parts of this work were completed.

\hspace*{\fill}

\bibliographystyle{unsrt}

\end{document}